\documentclass[
  reprint,
  superscriptaddress,
  preprintnumbers,
  amsmath,amssymb,
  aps,
  prl
]{revtex4-2}

\usepackage{graphicx}
\usepackage{dcolumn}
\usepackage{bm}
\usepackage{xcolor}
\usepackage{physics}
\usepackage{multirow}
\usepackage{orcidlink}
\usepackage{pgfplots}
\usepackage[normalem]{ulem}
\usepackage{nicefrac}
\usepackage{float}
\usepackage{tikz}
\usetikzlibrary{positioning}
\usepackage{comment}
\usepackage{booktabs}

\definecolor{siteAcolor}{HTML}{433247}   
\definecolor{siteBcolor}{HTML}{C79E5D}   
\definecolor{bondcolor}{HTML}{C4BEC4}    
\definecolor{pbccolor}{HTML}{C4BEC4}     

\newcommand{\latticescale}{0.55}
\newcommand{\latticesitesize}{2.2mm}

\tikzset{
  lattice/.style={
    scale=\latticescale,
    baseline=(current bounding box.center),
  },
  lsitestyle/.style={
    circle, draw=none, inner sep=0pt, outer sep=0pt,
    minimum size=\latticesitesize,
  },
  siteA/.style={lsitestyle, fill=siteAcolor},
  siteB/.style={lsitestyle, fill=siteBcolor},
  lbondstyle/.style={draw=bondcolor, line width=1.0pt, line cap=round},
  pbondstyle/.style={draw=pbccolor,  line width=0.7pt, line cap=round,
                     dash pattern=on 1.3pt off 1.5pt},
  panellabel/.style={font=\footnotesize, inner sep=0pt, anchor=base west},
}

\newcommand{\lsite}[3]{\node[#3] at (#1,#2) {};}
\newcommand{\lbond}[4]{\draw[lbondstyle] (#1,#2) -- (#3,#4);}
\newcommand{\pbond}[4]{\draw[pbondstyle] (#1,#2) -- (#3,#4);}
\newcommand{\lpanellabel}[3]{\node[panellabel] at (#1,#2) {#3};}

\newcommand{\hiskp}{Helmholtz Institute for Radiation and Nuclear Physics (HISKP), University of Bonn, Bonn, Germany}
\newcommand{\tra}{Transdisciplinary Research Area (TRA) Matter, University of Bonn, Bonn, Germany}
\newcommand{\ias}{Institute for Advanced Simulation 4 (IAS-4), Forschungszentrum J{\"u}lich, J{\"u}lich, Germany}

\begin{document}

\title{Tackling the Sign Problem in the Doped Hubbard Model with Normalizing Flows}
\author{Dominic Schuh\orcidlink{0000-0003-0866-1404}}\email{schuh@hiskp.uni-bonn.de}
    \affiliation{\tra}
    \affiliation{\hiskp}

\author{Lena Funcke\orcidlink{0000-0001-5022-9506}}
    \affiliation{\tra}
    \affiliation{\hiskp}

\author{Janik Kreit\orcidlink{0009-0000-2454-2339}}
    \affiliation{\tra}
    \affiliation{\hiskp}

\author{Thomas Luu\orcidlink{0000-0002-1119-8978}}%
    \affiliation{\hiskp}
    \affiliation{\ias}
    
\author{Simran Singh\orcidlink{0000-0002-9333-3925}}
    \affiliation{\tra}
    \affiliation{\hiskp}

\date{\today}

\begin{abstract}
\noindent
The Hubbard model at finite chemical potential is a cornerstone for understanding doped correlated systems, but simulations are severely limited by the sign problem. In the auxiliary-field formulation, the spin basis mitigates the sign problem, yet severe ergodicity issues have limited its use. We extend recent advances with normalizing flows at half-filling to finite chemical potential by introducing an annealing scheme enabling ergodic sampling. Compared to state-of-the-art hybrid Monte Carlo in the charge basis, our approach accurately reproduces exact diagonalization results while reducing statistical uncertainties by an order of magnitude, opening a new path for simulations of doped correlated systems.
\end{abstract}
\maketitle
\emph{Introduction}---Understanding systems of strongly correlated electrons is fundamental to quantum many-body-physics~\cite{Arovas_2022}. One of the most widely used theoretical frameworks for studying such systems is the Hubbard model~\cite{hubbard,hubbard1964electronII,hubbard1964electronIII}, which captures the essential competition between electron kinetic energy and on-site interactions, thereby giving rise to phenomena such as the Mott insulating behavior \cite{NFMott_1970}, magnetism and superconductivity~\cite{Qin:2021thl,PhysRevX.13.011007,doi:10.1126/science.adh7691}. 
Over the years, a wide range of methods have been developed to analyze the Hubbard model. 
In the weak-interaction regime, perturbative approaches provide valuable insights~\cite{Razmadze2024-mi}. In contrast, in the strongly correlated regime, where perturbative methods fail, non-perturbative methods are required. These can be broadly classified into Hamiltonian-based approaches and action-based approaches. The latter typically rely on the finite-temperature auxiliary-field formulation of the Hubbard model, where Monte Carlo simulations become essential (see, e.g., Refs.~\cite{Ostmeyer2020-od,Ostmeyer2024-ml,Sinilkov2025-xb,Rodekamp2025-wm,PhysRevB.98.235129,Beyl:2017kwp,Smith_2014,Ulybyshev_2013} and references therein).

Doped systems are described by a finite chemical potential and are essential for describing realistic materials and molecular systems. However, away from half-filling or on non-bipartite lattices, the Hubbard model suffers from a severe numerical sign problem, making reliable estimates of observables increasingly challenging \cite{Loh:1990zz}.  Furthermore, due to practical ergodicity problems in the spin basis, Monte Carlo based methods are usually restricted to work in the charge basis of the Hubbard model, where the sign problem is particularly pronounced. Although several methods have been developed to mitigate these challenges \cite{Rodekamp:2022xpf, Gantgen:2023byf}, they still fall short for crucial regions of the parameter space.

In this work, we build on the results of Refs. \cite{Schuh2025-ef, Schuh:2025gks} and develop the first deep generative machine learning approach to the finite-temperature auxiliary-field formulation of the Hubbard model at \textit{non-zero} chemical potential. 

To this end, we introduce an annealing scheme to overcome the practical ergodicity problem in the spin basis \cite{Wynen:2018ryx} and benchmark our method against results obtained by state-of-the-art optimized hybrid Monte Carlo (HMC) simulations \cite{Gantgen:2023byf}.

The remainder of this letter is organized as follows. We first introduce the Hubbard model in the spin basis and highlight its key features, followed by a brief overview of normalizing flows. We then present our proposed annealing scheme and its results, before concluding.

\emph{The Hubbard model at finite density}---The Hubbard model~\cite{hubbard,hubbard1964electronII,hubbard1964electronIII,brower2012hybridmontecarlosimulation} provides a simple framework for studying the interaction of electrons in materials and thereby their conducting properties. In its simplest form, the dynamics consist of nearest-neighbor hopping, a repulsive Coulomb-like on-site interaction, and a chemical potential. In this work, we will focus on studying the model in the spin basis and at finite chemical potential. The corresponding Hamiltonian can be written as
\begin{equation}\label{eqn:hubbard}
    H = -\kappa\sum_{\langle x,y \rangle,\sigma} c_{x,\sigma}^\dagger c_{y,\sigma} - \frac{U}{2} \sum_x \left( n_{x,\uparrow}-n_{x,\downarrow} \right)^2
    - \mu \sum_x  q_x \, ,
\end{equation}
where $\kappa$ is the hopping parameter, $U$ the on-site interaction, and $\mu$ the chemical potential. Here, $c_{x,\sigma}^\dagger \, (c_{x,\sigma})$ denotes the creation (annihilation) operator for an electron at site $x$ with spin $\sigma\,(=\uparrow\downarrow)$, $n_{x, \sigma} = c_{x,\sigma}^\dagger c_{x,\sigma}$ is the corresponding number operator, and $q_x = \sum_{\sigma} n_{x,\sigma}$ is the net charge at site $x$ \footnote{Note that Eq.~\eqref{eqn:hubbard} is equivalent to the standard form of the repulsive-Hubbard interaction featuring the term $U \sum  n_{x,\uparrow}  n_{x,\downarrow}$, with a shifted chemical potential $\bar{\mu} = \mu + \frac{U}{2}$, up to an irrelevant constant.}.

When put on a honeycomb lattice with periodic boundaries, as displayed in Fig.~\ref{fig:hex_lattices}, the Hubbard model is known to describe carbon-based materials, such as graphene \cite{Ostmeyer2020-od}, quite well. 

\begin{figure}
    \centering
%
\begin{tikzpicture}[lattice]
  \path (0,-2.40) (0,2.55);
  \lpanellabel{-2.7}{2.55}{(a)}

  \lbond{1.250}{0.000}{0.625}{1.083}
  \lbond{0.625}{1.083}{-0.625}{1.083}
  \lbond{-0.625}{1.083}{-1.250}{0.000}
  \lbond{-1.250}{0.000}{-0.625}{-1.083}
  \lbond{-0.625}{-1.083}{0.625}{-1.083}
  \lbond{0.625}{-1.083}{1.250}{0.000}
  \lbond{-1.250}{0.000}{-2.500}{0.000}
  \lbond{1.250}{0.000}{2.500}{0.000}
  \pbond{-2.500}{0.000}{0.625}{1.083}
  \pbond{-2.500}{0.000}{0.625}{-1.083}
  \pbond{2.500}{0.000}{-0.625}{1.083}
  \pbond{2.500}{0.000}{-0.625}{-1.083}
  \lsite{1.250}{0.000}{siteA}
  \lsite{-0.625}{1.083}{siteA}
  \lsite{-0.625}{-1.083}{siteA}
  \lsite{-2.500}{0.000}{siteA}
  \lsite{0.625}{1.083}{siteB}
  \lsite{-1.250}{0.000}{siteB}
  \lsite{0.625}{-1.083}{siteB}
  \lsite{2.500}{0.000}{siteB}
\end{tikzpicture}
\hfill
%
\begin{tikzpicture}[lattice]
  \path (0,-2.40) (0,2.55);
  \lpanellabel{-4.575}{2.55}{(b)}

  \lbond{-4.375}{0.000}{-3.125}{0.000}
  \lbond{-2.500}{1.083}{-1.250}{1.083}
  \lbond{-0.625}{2.166}{0.625}{2.166}
  \lbond{-2.500}{-1.083}{-1.250}{-1.083}
  \lbond{-0.625}{0.000}{0.625}{0.000}
  \lbond{1.250}{1.083}{2.500}{1.083}
  \lbond{-0.625}{-2.166}{0.625}{-2.166}
  \lbond{1.250}{-1.083}{2.500}{-1.083}
  \lbond{3.125}{0.000}{4.375}{0.000}
  \lbond{-2.500}{1.083}{-3.125}{0.000}
  \lbond{-0.625}{2.166}{-1.250}{1.083}
  \lbond{-2.500}{-1.083}{-3.125}{0.000}
  \lbond{-0.625}{0.000}{-1.250}{1.083}
  \lbond{-0.625}{0.000}{-1.250}{-1.083}
  \lbond{1.250}{1.083}{0.625}{2.166}
  \lbond{1.250}{1.083}{0.625}{0.000}
  \lbond{-0.625}{-2.166}{-1.250}{-1.083}
  \lbond{1.250}{-1.083}{0.625}{0.000}
  \lbond{1.250}{-1.083}{0.625}{-2.166}
  \lbond{3.125}{0.000}{2.500}{1.083}
  \lbond{3.125}{0.000}{2.500}{-1.083}
  \pbond{-4.375}{0.000}{0.625}{2.166}
  \pbond{-4.375}{0.000}{0.625}{-2.166}
  \pbond{-2.500}{1.083}{2.500}{-1.083}
  \pbond{-2.500}{-1.083}{2.500}{1.083}
  \pbond{-0.625}{2.166}{4.375}{0.000}
  \pbond{-0.625}{-2.166}{4.375}{0.000}
  \lsite{-4.375}{0.000}{siteA}
  \lsite{-2.500}{1.083}{siteA}
  \lsite{-2.500}{-1.083}{siteA}
  \lsite{-0.625}{2.166}{siteA}
  \lsite{-0.625}{0.000}{siteA}
  \lsite{-0.625}{-2.166}{siteA}
  \lsite{1.250}{1.083}{siteA}
  \lsite{1.250}{-1.083}{siteA}
  \lsite{3.125}{0.000}{siteA}
  \lsite{-3.125}{0.000}{siteB}
  \lsite{-1.250}{1.083}{siteB}
  \lsite{-1.250}{-1.083}{siteB}
  \lsite{0.625}{2.166}{siteB}
  \lsite{0.625}{0.000}{siteB}
  \lsite{0.625}{-2.166}{siteB}
  \lsite{2.500}{1.083}{siteB}
  \lsite{2.500}{-1.083}{siteB}
  \lsite{4.375}{0.000}{siteB}
\end{tikzpicture}
    \caption{The two honeycomb lattices used in this work, with (a) $2\times2$ ($N_x=8$ sites) and (b) $3\times3$ ($N_x=18$ sites) unit cells. Dark and light circles denote the two triangular sublattices, solid lines the nearest-neighbour bonds, and dashed lines the bonds reflecting the periodic boundaries.}
    \label{fig:hex_lattices}
\end{figure}

However, with increasing system size, the Hilbert space grows exponentially with the number of spatial sites $N_x$, rendering exact methods intractable. To circumvent this problem, the model can be reformulated in an action-based auxiliary field representation \cite{hubbard,hubbard1964electronII,hubbard1964electronIII,brower2012hybridmontecarlosimulation,Luu_2016,rodekamp-phd}, enabling the use of Monte-Carlo based methods. In this framework, observables are expressed as statistical quantum-mechanical expectation values,
\begin{equation}
    \langle \mathcal{O} \rangle = \frac{1}{\mathcal{Z}} \mathrm{tr} \left\{e^{-\beta H} \mathcal{O}\right\} \,,
\end{equation}
where the partition function is $\mathcal{Z} = \mathrm{tr}\left\{e^{-\beta H}\right\}$, the inverse temperature is $\beta=1/k_B T$, and the trace runs over the entire Fock space. 

After applying a Suzuki-Trotter decomposition \cite{trotter1959product, 10.1143/PTP.56.1454} and a Hubbard-Stratonovich transformation \cite{Hubbard:1959ub} to decouple the quartic fermion interaction, this expectation value can be rewritten in terms of an auxiliary bosonic field $\phi$,
\begin{equation}
    \langle \mathcal{O} \rangle = \frac{1}{\mathcal{Z}} \int \mathcal{D}[\phi] e^{-S[\phi]} \mathcal{O}[\phi].
\end{equation}
Here, the Hubbard action $S[\phi]$ is given by 
\begin{equation}
    S[\phi] = \frac{1}{2\tilde{U}} \sum_{x,t \in \Lambda} \phi_{x,t}^2 
     - \ln\det \left(M[\phi|\tilde{\kappa},\tilde{\mu}]M[-\phi|\tilde{\kappa},\tilde{\mu}]\right),
     \label{eq:S}
\end{equation}
where $\Lambda = N_x \times N_t$ is the lattice volume, with $N_x$ denoting the number of physical lattice sites and $N_t$ the number of discrete time slices, $\phi_{xt}$ is the field at site $(x,t)$, and $\tilde{U}=U\beta/N_t$, $\tilde{\kappa}=\kappa \beta/N_t$, $\tilde{\mu}=\mu \beta/N_t$ are the dimensionless lattice couplings. The fermion matrices $M[\pm \phi, \tilde{\kappa}, \tilde{\mu}]$ originate from the spin-up and spin-down electrons in Eq.~\eqref{eqn:hubbard}, respectively.

In this reformulation, one can choose a discretization of the fermion matrix \cite{Ulybyshev_2013}. In this work, we adopt the exponential discretization, as it preserves chiral (or sub-lattice) symmetry \cite{PhysRevB.98.235129,Wynen:2018ryx}. In this discretization, the matrix elements are given by
\begin{equation}\label{eqn:matel}
    M[\phi|\tilde{\kappa}, \tilde{\mu}]_{x't',xt} = \delta_{x',x} \delta_{t' t} - [e^h]_{x'x} e^{\phi_{xt} + \tilde{\mu}} \mathcal{B}_{t'} \delta_{t',t+1}\,.
\end{equation}
Here, $h$ is the hopping matrix defined as $h = \tilde{\kappa}\delta_{\langle x,x' \rangle}$, with $\langle x,x' \rangle$ denoting nearest-neighbor-interactions, and $\mathcal{B}$ incorporates anti-periodic boundary conditions in the temporal direction \footnote{Note that we could have equivalently defined the chemical potential term on the diagonal of the hopping matrix}. 

The observable considered in this work is the single-particle Euclidean-time correlator, defined as
\begin{equation}\label{Eq:Ct}
    C_{x,y}(t) \equiv \langle c_x(t) c^{\dagger}_y(0) \rangle = \frac{1}{N_{\text{cfg}}}\sum_{\{\phi\}} M^{-1}_{(t,x);(0,y)} [\phi|\tilde{\kappa},\tilde{\mu}].
\end{equation}
These correlators are computed for all site pairs $(x,y)$ on $N_{\text{cfg}}$ configurations and projected onto the eigenvectors of the hopping matrix $h$, yielding $N_x$ eigen-correlators. For a bipartite lattice with $N_x/2$ unit cells, each containing one site of either sublattice, this projection is equivalent to a spatial Fourier transform over unit cells followed by diagonalization of the resulting $2\times2$ sublattice block at each momentum $\mathbf{k}$~\cite{Wynen:2018ryx},
\begin{equation}\label{Eq:Ck}
    C_{\pm}(\mathbf{k},t) = \frac{1}{N_{\text{cell}}}
    \sum_{\mathbf{x}} e^{i\mathbf{k}\cdot\mathbf{x}}\,
    C_{\pm}(\mathbf{x},t),
\end{equation}
where the sum runs over unit-cell positions $\mathbf{x}$ and the $\pm$ index labels the symmetric and antisymmetric sublattice combinations. This observable has previously been used to illustrate the lack of ergodicity in sampling the Boltzmann distribution \cite{Wynen:2018ryx,Schuh:2025gks}. In this work, we demonstrate the ergodic nature of our novel annealing-based sampling method by comparing these correlators with results obtained from exact diagonalization (ED), where available.

\emph{Choice of basis}---\noindent The Hubbard model is most commonly simulated either in the repulsive spin basis \cite{Gottlieb:1987mq,Beyl:2017kwp} or the attractive charge basis \cite{Ulybyshev_2013, Luu_2016}. In the action-based formulation, the difference between the two lies in how the auxiliary field $\phi_{xt}$ enters the fermion matrix in Eq. \ref{eqn:matel}. In the spin basis the fields are real-valued, making the matrix elements real, whereas in the charge basis, the fields are purely imaginary. 

Monte Carlo simulations of the Hubbard model typically face two major challenges: (i) ergodicity issues~\cite{Beyl:2017kwp,Wynen:2018ryx} and (ii) the non-positivity of the probability weight arising from the product of fermion determinants~\cite{Scalettar:1989ig,Iglovikov_2015}. Consequently, at zero chemical potential, the feasibility of HMC simulations depends primarily on effectively addressing the ergodicity problem. This aspect has been systematically investigated in Refs.~\cite{Wynen:2018ryx, temmen2024overcomingergodicityproblemshybrid}, where the authors show that the formal ergodicity issues of the charge basis are often easier to handle than the practical ergodicity issues of the spin basis. 

Since the primary objective of this work is to investigate the system at finite chemical potential, it is advantageous to adopt a formulation that offers better control over the sign problem. In this context, as we explicitly demonstrate, the spin basis offers a structural advantage: the fermion matrix is real for each spin species separately, thereby avoiding the complex probability weights that arise in the charge basis. As a result, the sign problem is purely real rather than complex, which simplifies its numerical treatment.

In the spin basis, the fermion matrices of the two species are real but not related by conjugation, so the associated probability weight is not manifestly positive. This can lead to a potential sign problem even at half-filling. However, as shown in Ref.~\cite{Wynen:2018ryx}, the fermion determinant in the exponential discretization satisfies a symmetry relation between the two species that guarantees positivity of the probability weight at zero chemical potential. Introducing a finite chemical potential explicitly breaks this symmetry and thus destroys the exact positivity of the measure \footnote{The relation between fermion matrices at finite chemical potential and the emergence of a sign problem is derived in Ref.~\cite{SupMat}.}.

\emph{Sign quenching \& re-weighting---}When simulating theories with a complex action, standard sampling-based methods — which rely on a probabilistic interpretation of the Boltzmann weight \( e^{-\beta S} \) — are no longer directly applicable. In such cases, one typically follows one of two strategies: (i) simulate the \emph{phase-quenched} theory \cite{Fodor:2001au}, in which the complex phase of the Boltzmann weight is neglected and subsequently restored by reweighting; or (ii) simulate the \emph{sign-quenched} theory \cite{deForcrand:2002pa, Giordano:2020roi}, where the Boltzmann weight is replaced by the absolute value of its real part, 
and sign fluctuations are reintroduced through \emph{sign reweighting}. The phase-quenched approach is primarily hindered by severe overlap problems \cite{Fodor:2001pe}, whereas the sign-quenched approach suffers from an exponentially degrading signal-to-noise ratio \cite{Troyer:2004ge,Aarts:2015tyj}. 

In the spin basis, the sign problem is \emph{literally} a sign problem: the probability weight is real but not positive definite, rather than complex. In order to explain the sign-reweighting procedure, we consider a theory described by a Boltzmann weight $w_t$. We therefore employ sign quenching, defining $w_q = |w_t|$ for real weights $w_t$ that fluctuate in sign. Observables in the full theory, $\langle O \rangle_t$, are recovered via reweighting:
\begin{align}\label{eq1:srweight}
    \langle O \rangle_t &= \frac{\int \mathcal{D} \phi \,\, w_t(\phi) O(\phi)}{\int \mathcal{D} \phi \,\, w_t(\phi)} 
    = \frac{\langle \frac{w_t}{w_q}\cdot O \rangle_q}{\langle \frac{w_t}{w_q} \rangle_q} \,,
\end{align}
where $\langle\cdot\rangle_q$ denotes sampling from the sign-quenched theory.

The severity of the sign problem is quantified by the denominator in Eq.~\eqref{eq1:srweight}, commonly referred to as the \textit{average sign} $\Sigma$. A value of $\Sigma = 1$ corresponds to the absence of a sign problem, whereas $\Sigma \to 0$ indicates the most severe sign problem, where cancellations between configurations suppress the signal. 
The resulting reduction in statistical efficiency leads to an effective sample size $N^{\mathrm{eff}} = |\Sigma|^2 N$ \cite{BERGER20211}, with Monte Carlo uncertainties scaling as $(N^{\mathrm{eff}})^{-1/2}$. 

With this in mind, we model the \textit{sign-quenched} distribution $e^{-\beta S_{q}}$, where
\begin{equation}
    S_{q}[\phi] = \sum\limits_{x,t \in \Lambda} \frac{\phi_{x,t}^2}{2\tilde{U}}
     -  \log | \det \left( M [\phi|\tilde{\kappa}, \tilde{\mu}] M[-\phi|\tilde{\kappa},\tilde{\mu}] \right) |\,.
\end{equation}
This gives us a well-defined probability weight, from which the physical theory can be recovered by reweighting observables with the sign of the original weight. 

\begin{figure*}[t]
    \centering
    \includegraphics[width=0.9\textwidth]{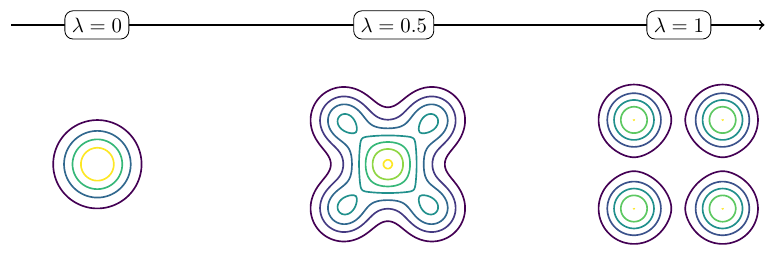}
    \caption{Schematic illustration of the annealing procedure controlled by the parameter $\lambda$. At $\lambda=0$, the fermion determinant is absent and the target distribution is Gaussian. As $\lambda$ is increased, fermionic effects are gradually introduced, deforming the distribution until the full interacting Hubbard action with its multimodal structure is recovered at $\lambda=1$.}
    \label{fig:annealing}
\end{figure*}

\emph{Normalizing flows}---
Normalizing flows (NFs) \cite{kobyzev2020normalizing,nfreview} are a class of generative models that enable efficient sampling from high-dimensional distributions. This is achieved by learning a bijective map $f_\theta$ between a simple prior distribution $q_z$ and a target distribution $q_y$, 
\begin{equation}
    f_\theta: z \sim q_z \mapsto y \sim q_y \,,
\end{equation}
where $f_\theta$ is parametrized by neural network weights $\theta$. In our setup, the target degrees of freedom $y$ are the auxiliary field configurations of the Hubbard model. Constructing these maps requires balancing the expressivity of the neural network with the need for a tractable Jacobian. Several approaches have been developed to this end, such as coupling-based NFs \cite{nice,realnvp,splineflows}, autoregressive NFs \cite{autoregressive}, and continuous NFs \cite{continuousflow}.

In this work, we employ a coupling-based NF with a real-valued non-volume preserving (RealNVP) architecture~\cite{realnvp}. Specifically, we consider a sequence of $L$ smooth, invertible mappings $f^l_{\theta_l}: \mathbb{R}^{|\Lambda|} \rightarrow \mathbb{R}^{|\Lambda|}$, each defined by a neural network with parameters $\theta_l$. Since each layer is bijective, their composition defines an overall invertible transformation:
\begin{equation}
y \equiv f_\theta(z) = \left( f^L_{\theta_L} \circ f^{L-1}_{\theta_{L-1}} \circ \cdots \circ f^1_{\theta_1} \right)(z) \, ,
\end{equation}
where $\theta = \{\theta_1, \ldots, \theta_L\}$ denotes the full set of trainable parameters. A crucial property of this construction is that the determinant of the Jacobian of $f_\theta$ can be computed efficiently, which is required for tractable likelihood-based training and density estimation of $q_y$. 

The generated field configurations $\phi$ are then given by
\begin{equation}
    \phi \equiv f_\theta (z) \,,
\end{equation}
with the associated learned density
\begin{equation}
    q_\theta(f_\theta(z)) = q_z(z) \cdot \left| \frac{\partial f_\theta}{\partial z} \right|^{-1} \,.
\end{equation}
Training of the neural network proceeds by minimizing the reverse Kullback–Leibler (KL) divergence \cite{kullback1951information}, which measures the distance between the flow-induced distribution $q_\theta$ and the target Boltzmann distribution $p(\phi) = e^{-S[\phi]}/Z$. This objective can be written as
\begin{equation}\label{eq:self_reg_kl}
\mathrm{KL}(q_\theta||p) = \mathbb{E}_{z \sim q_z} \Big[\log q_\theta\left(f_\theta(z)\right) + S\left[f_\theta(z)\right] \Big] \,,
\end{equation}
where $S[\cdot]$ denotes the action of the underlying physical system. To ensure exact and unbiased sampling, observables calculated from the flow are subsequently reweighted according to the unnormalized importance weights $w(\phi) = \exp (-S[\phi])/ q_\theta(\phi)$, where, in our case, $S=S_q[\phi]$ corresponds to the sign-quenched action.

\emph{Avoiding practical ergodicity problems with normalizing flows}---It is well known that training with the reverse KL divergence can lead to mode dropping in the learned density \cite{Minka2005Divergence,Hackett:2021idh,Nicoli:2023qsl}. In challenging settings, such as the Hubbard model in the spin basis, having an ergodic sampler is essential to ensure unbiased results \cite{Wynen:2018ryx, Schuh:2025gks}. Symmetry-informed architectures, e.g. canonicalization \cite{Schuh2025-ef} or symmetry-enforcing stochastic modulation \cite{sesamo,Schuh:2025gks}, have proven highly effective at sampling distributions with a strong practical ergodicity problem. However, their applicability is limited to systems exhibiting particular symmetries, which in the Hubbard model typically arise only in the strongly coupled regime, where these symmetries become particularly pronounced.

To describe systems across a range of couplings, we cannot rely on deterministic symmetry transformations. Instead, we employ an annealing scheme~\cite{annealing}, in which an auxiliary parameter $\lambda$ scales the fermionic contribution to the action,
\begin{equation}
    S_{q,\lambda}[\phi] = \sum\limits_{x,t \in \Lambda} \frac{\phi_{x,t}^2}{2\tilde{U}}
     - \lambda \cdot  \log \left|\det \left(M [\phi|\tilde{\kappa}, \tilde{\mu}] M[-\phi|\tilde{\kappa},\tilde{\mu}] \right) \right| .
\end{equation}
For $\lambda = 0$, the fermion determinant is absent, leaving a simple Gaussian distribution, while $\lambda = 1$ recovers the full interacting Hubbard action. Gradually increasing $\lambda$ during training smoothly deforms the target distribution toward the final multimodal target, as illustrated schematically in Fig.~\ref{fig:annealing}. This mitigates the mode-seeking behavior of the reverse KL divergence, allowing the NF to receive training signals from all modes and represent even widely separated probability sectors. 

This approach is highly scalable, as its computational cost grows only moderately with system size, adding a minimal training overhead. We emphasize, however, that while the annealing procedure empirically demonstrates strong robustness against mode dropping, it does not provide a formal guarantee of complete mode coverage in the final model. 

\emph{Results}---In the following, we evaluate the performance of our annealing-based NF ansatz for the Hubbard model in the spin basis. As a benchmark, we use state-of-the-art HMC in the charge basis with optimized shifts \cite{Gantgen:2023byf}. We compare the average sign of both approaches and calculate correlation functions for the most challenging system. Each model is trained for 50k steps, with the annealing parameter $\lambda$ increased linearly from 0 to 1 during the first 1k steps. 

\begin{figure}[h]
    \centering
    \includegraphics[width=0.95\linewidth]{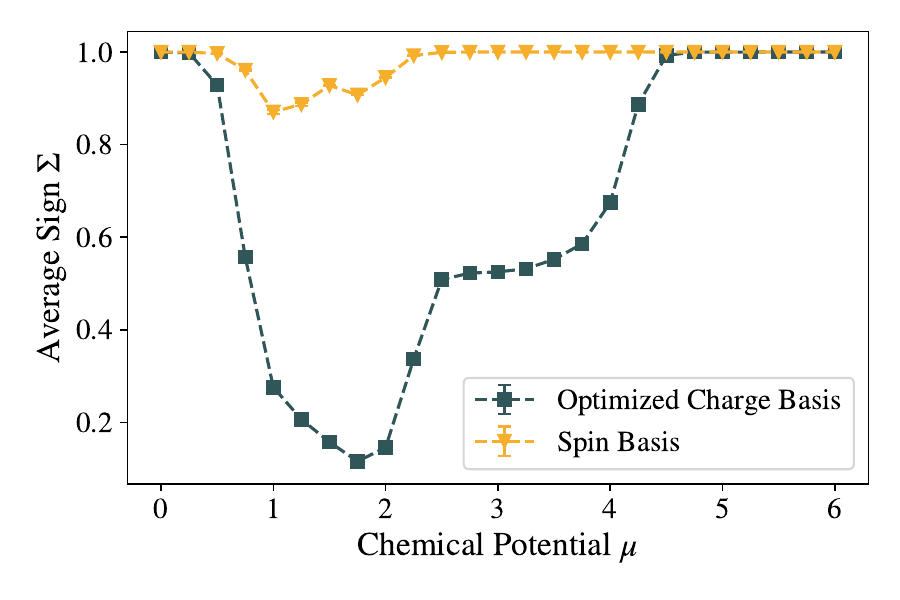}
    \caption{Average sign $\Sigma$ as a function of the chemical potential $\mu$ for an eight-site hexagonal lattice with $U = 2$, $\beta = 8$, and $N_t = 16$. Shown are results from the annealing-based NF approach in the spin basis (yellow) and optimized HMC simulations in the charge basis~\cite{Gantgen:2023byf} (grey). Uncertainties are shown but too small to be visible.}
    \label{fig:avg_sign}
\end{figure}

Figure~\ref{fig:avg_sign} shows the average sign obtained with the NF ansatz compared to optimized HMC simulations for an eight-site hexagonal lattice with $U = 2$, $\beta = 8$, and $N_t = 16$. Although optimized shifts are known to alleviate the sign problem in the charge basis compared to standard HMC~\cite{Gantgen:2023byf}, we still observe a significant suppression of the average sign for $\mu \in [1.0, 2.0]$ in this setup. In contrast, simulations performed in the spin basis using the NF approach maintain a substantially larger average sign across this region, with an improvement of approximately a factor of four at the minimum.

\begin{figure}[b]
    \centering
    \includegraphics[width=0.95\linewidth]{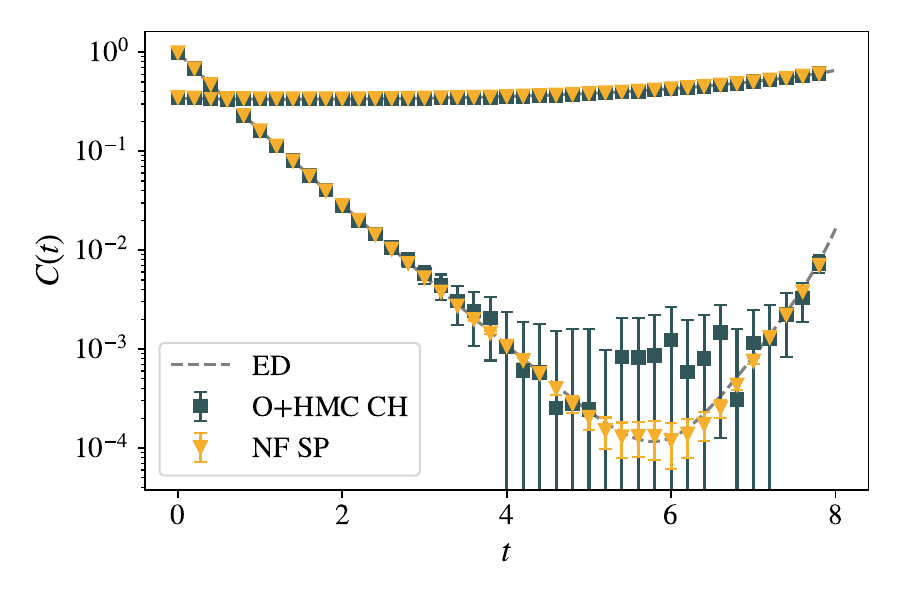}
    \caption{One-body correlation functions $C_+((0,0),t)$ (upper  curve) and $C_+((\pi/3,-\pi/\sqrt{3}),t)$ (lower curve) for an eight-site hexagonal lattice with $U=2$, $\beta=8$, $\mu=1.75$, and $N_t=40$, corresponding to an electron doped system with doping $\delta=-0.49(1)$. Shown are continuous exact diagonalization results (ED, grey dashed), state-of-the-art optimized HMC in the charge basis (O+HMC CH, grey), and our annealing-based NF ansatz in the spin basis (NF SP, yellow). Both HMC and NF results are generated with 1M samples each.}
    \label{fig:8sites_corr}
\end{figure}

To validate and assess our approach, we calculate one-body correlation functions $C_\pm(\bm{k},t)$, defined in Eq.~\ref{Eq:Ck}, and compare them to the ground truth obtained via ED as well as the state-of-the-art optimized HMC algorithm. We consider an eight-site hexagonal lattice with parameters $U=2$, $\beta=8$, $\mu=1.75$, and $N_t=40$, which represents the largest exactly solvable system exhibiting the most severe sign problem, as indicated in Fig.~\ref{fig:avg_sign}. The corresponding results are shown in Fig.~\ref{fig:8sites_corr}. The two correlators shown here serve different purposes. The upper branch is a consistency check rather than a demonstration of improvement. The improvement appears in the lower branch, which is the observable used in Refs.~\cite{Schuh:2025gks, Wynen:2018ryx} to expose non-ergodic sampling.

We find that the NF approach reproduces the exact results with high precision while reducing uncertainties by an order of magnitude compared to HMC. Additional results for $U=3$ are provided in Ref.~\cite{SupMat}. 

\begin{figure}[t]
    \centering
    \includegraphics[width=0.95\linewidth]{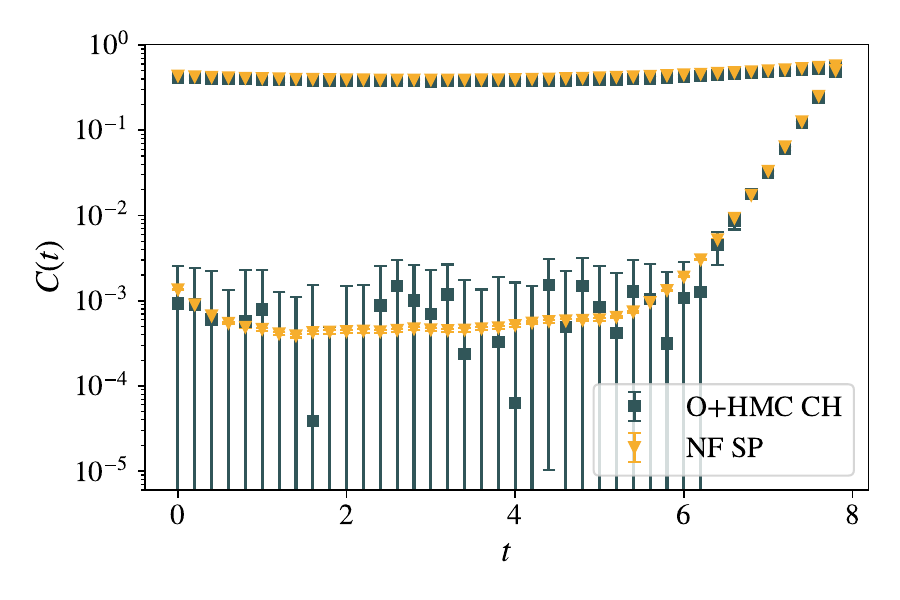}
    \caption{One-body correlation functions $C_+((-4\pi/9,0),t)$ (upper curve) and $C_-((-4\pi/9,0),t)$ (lower curve) for an 18-site hexagonal lattice with $U=2$, $\beta=8$, $\mu=2.5$, and $N_t=40$, corresponding to an electron doped system with doping $\delta=-0.50(1)$. Shown are state-of-the-art optimized HMC in the charge basis (O+HMC CH, grey) and our annealing-based NF ansatz in the spin basis (NF SP, yellow). Both HMC and NF results are generated with 1M samples each.}
    \label{fig:18sites_corr}
\end{figure}

Lastly, we extend our analysis beyond exactly solvable systems and compute one-body correlation functions for an 18-site hexagonal lattice with parameters $U=2, \beta=8, \mu=2.5,$ and $N_t=40$, as shown in Fig.~\ref{fig:18sites_corr}. In the absence of exact results at this system size, we validate our approach using a less noise-sensitive correlator, finding good agreement with HMC (top correlator in Fig.~\ref{fig:18sites_corr}). For more noise-sensitive observables (bottom correlator), the NF method produces smaller uncertainties and improved accuracy, consistent with the milder sign problem in the spin-basis formulation.

\emph{Conclusion}---In this work, we presented a new annealing-based NF method that enables reliable sampling of the Hubbard model in the spin basis at \textit{finite} chemical potential. By learning the sign-quenched theory and exploiting the milder sign problem in this formulation, our approach outperforms state-of-the-art optimized HMC calculations by up to an order of magnitude in both accuracy and uncertainty for system sizes up to $N_x = 18$. 

While we have focused on hexagonal lattices throughout this work in order to allow for comparison with existing work, the annealing scheme makes no
use of the lattice structure itself, and we expect its effectiveness to carry over to other
geometries. The same applies to the reality of the fermion matrix underlying the milder sign
problem in the spin basis, although on non-bipartite lattices the positivity of the weight is lost
and reweighting would already be required at zero chemical potential. The severity of the resulting
sign problem in that case remains an open question for future investigation.

The annealing scheme enables controlled sampling of multimodal target distributions without introducing significant training overhead or strong dependence on the system size. As a result, the overall scalability of the method is primarily determined by the properties of the underlying generative model. While the RealNVP architecture employed here performs well at the lattice sizes considered, its training cost is expected to increase for larger volumes. 
Building on recent work on improving the numerical stability of action-based Hubbard model simulations~\cite{kreit2026scalablenormalizingflowshubbard}, we plan to extend our annealing scheme to more expressive and scalable generative models in future work.

Taken together, these results establish a broadly applicable framework for efficient and accurate sampling of the Hubbard model in regimes where conventional Monte Carlo methods are limited by ergodicity loss or severe sign problems.

\emph{Acknowledgements}---The authors thank Johann Ostmeyer for insightful discussions. The authors gratefully acknowledge the granted access to the Marvin cluster hosted by the University of Bonn, as well as the computing time
granted by the JARA Vergabegremium and provided on the JARA Partition part of the supercomputer JURECA at
Forschungszentrum Jülich~\cite{jureca-2021}. This project was supported by the Deutsche Forschungsgemeinschaft (DFG, German Research Foundation) as part of the CRC 1639 NuMeriQS – project no. 511713970.

\emph{Data Availability.}---The data that support the findings of this article are openly available \cite{data}.

\bibliography{refs}
\end{document}